\documentstyle[12pt]{article}
\begin{titlepage}   
\title{ 
{
\vspace{-3.5ex}
\normalsize\bf
\begin{flushright}
IHEP 99--15\\
\end{flushright}
\vspace{10ex}
}
{\bf Quark masses and mixings}\\
{\bf  in the standard model }\\
{\bf with heavy vector-like families}\\[3ex]
}

\author{ Yury F.~Pirogov{}\thanks{E-mail: pirogov@mx.ihep.su}
\and Oleg V.~Zenin \\[1ex]
{\it Institute for High Energy Physics,}\\
{\it Protvino, RU-142284 Moscow Region, Russia}\\[0.5ex]
{\it  Moscow  Institute  of  Physics  and  Technology,}\\
{\it  Dolgoprudny,  Moscow  Region, Russia }}

\date{}

\begin{document}
\maketitle
\thispagestyle{empty}

\abstract{
\noindent
The  extension of the standard model with  pairs of 
the vector-like families is studied. 
The quark mixing matrices for the  left- and right-handed charged
currents, 
as well as those for the flavour changing neutral currents, 
the $Z$ and Higgs mediated, are found. 
Both the model independent parametrization for an arbitrary case and
an
explicit realization for the case with one pair of the heavy
vector-like families are presented.
The extension opens new prospects for studying  deviations 
from the standard model in the future experiments at high energies.
}

\thispagestyle{empty}
\end{titlepage}

\addtocounter{page}{1}

\section{ Introduction }
At present we know of three quark-lepton chiral families in the
standard model (SM). Their mixing within the present
experimental accuracy is well known to be described by
the  $3\times 3$ unitary matrix~\cite{KM}. But beyond it, 
whether there are extra families and, if so, what  their masses and
mixings are --- this is yet unsolved problem. 

A recent two-loop renormalization group analysis~\cite{pir} of the SM
shows that subject to
the precision experiment restriction on the Higgs
mass, $M_H\le 215$~GeV at 95\%~C.L.~\cite{tournefier}, the forth
chiral family, if alone, is excluded.\footnote{The recent 
more conservative restrictions 
$m_H\le 262$ GeV or $M_H\le 300$ GeV at $95\%$ C.L., respectively,
from the first and second papers of Ref.~\cite{higgs} render the
fourth chiral family only marginally possible.}
In fact, it does not depend on whether
this extra family has the normal chiral structure or the mirror one. 
But as it  is noted  in  Ref.~\cite{pir},
a pair of the opposite chirality  families with the 
relatively low Yukawa couplings evades the SM self-consistency
restrictions and could still exist. In order to
conform to observations these extra families, which otherwise can
be considered as the vectorial ones,  should get large
direct masses and  drop out of the light particle spectrum of the
SM in the decoupling limit. Nevertheless, 
at the not too high masses, say, in the TeV region,
such families  could result in observable
corrections to the SM interactions through mixing with  the light
fermions. 

Various vector-like fermions are generic  in many extensions of
the SM like the superstring and grand unified theories, composite
models, etc. Many issues concerning those fermions, both the
electroweak doublets and singlets, the latter ones of the up and down
types, were considered in the literature~\cite{vlf},~\cite{lavoura}. 
On the other hand there are numerous studies of the $n>3$ chiral
family extensions of the SM~\cite{n_families}, \cite{santa}. 
Some topics concerning the SM extensions with the vector-like
families are studied in Ref.~\cite{fuji}.

In a previous letter \cite{VLF2} we presented the results for 
the SM light quark masses and mixings in the presence of
the extra vector-like families.
In the current paper we give the complete results including those for 
the heavy quarks. In Section 2 we carry out the model independent
analysis for the general case. In Section 3 an explicit realization
for the case with a pair of the heavy vector-like families is
presented. In  Appendix we give the technical details of the
diagonalization procedure and the explicit form of the mixing
matrices through the elements of the general mass matrices.

\section{Model independent analysis}

The most general content of the SM families  consisting of the
$\mathrm SU(2)_W \times U(1)_Y$  doublets and singlets  is
illustrated in Table~1. The notations with a hat sign designate
quarks in the symmetry/electroweak
basis where, by definition, the SM symmetry structure is well stated.
``Normal'' in the row means  the $n\ge 3$ chiral families, similar in
their chiral and quantum number
structure to three ordinary families of the minimal SM. ``Mirror''
means  the $m\ge 0$ mirror conjugate families with the normal quantum
numbers, or in other terms, the charge conjugate families with the
normal chiral structure.  We suppose for definiteness that $n>m$.
``Chiral'' in the column means  the  chiral notations, and
``mixed'' corresponds to  the more traditional  
left-right notations.\footnote{To be as clear as possible, 
what we are talking about, say, in terms of the 15-plets of the GUT
$SU(5)$ ($15 = 10\oplus \overline{5}$)
is $n 15_L\oplus m 15_R$, or $n 15_L\oplus m \overline{15}_L$.
Nevertheless, the scales we have in mind are much lower than those 
of the GUT's, typically ${\cal O} (1 - 100$) TeV, i.e.\ rather those
of the composite models.}

\begin{table}[htbp]
\paragraph{Table 1} The general content of the SM families.

\vspace{1ex}
\begin{tabular}{|c|c|c|c|}
\hline
&$\#$&Chiral&Mixed\\
\hline
Normal&$n$&$Q_L=({\hat q}_L,{\hat u}_L^c, {\hat d}_L^c)$&$({\hat
q}_L,
{\hat u}_R, {\hat d}_R)$\\
Mirror&$m$&$Q'_R=({\hat q}_R', {\hat u}'^c_R, {\hat d}'^c_R)$&$({\hat
q}'_R, {\hat u}'_L, {\hat d}'_L)$\\
\hline
\end{tabular}
\end{table}

In general, quarks gain masses from two different physical
mechanisms: that of the SM Yukawa interactions and that of a New
Physics resulting in the SM invariant direct mass terms.
Being chirally unprotected the latter ones should naturally 
be characterized by a high mass scale $M$, $M\gg v$, with $v$
being the SM Higgs vacuum expectation value.
In the symmetry basis the kinetic, Yukawa and direct mass
Lagrangian has the following most general form:
\def\D{D\hspace{-0.28cm}/\hspace{0.08cm}}
\begin{eqnarray}
\label{eq:lagrangian}
{\cal L}&=&
~~~i \overline{{\hat q}_L}\D {\hat q}_L + i \overline{{\hat u}_R}\D
{\hat u}_R + 
i \overline{{\hat d}_R}\D {\hat d}_R\nonumber\\
&&+\, i\overline{{\hat q}'_R}\D {\hat q}'_R + i \overline{{\hat
u}'_L}\D {\hat u}'_L + 
i \overline{{\hat d}'_L}\D {\hat d}'_L\nonumber\\
&&-\Big{(}\overline{{\hat q}_L} Y^u {\hat u}_R \phi^c +
\overline{{\hat q}_L} Y^d {\hat d}_R\phi
+\overline{{\hat u}'_L} {Y^u}' {\hat q}'_R {\phi^c}^{\dagger} 
+ \overline{{\hat d}'_L} {Y^d}' {\hat q}'_R \phi^{\dagger} 
+ \mbox{h.c.}\Big{)}
\nonumber\\
&& - \Big{(}
 \overline{{\hat q}_L} M {\hat q}'_R + 
\overline{{\hat u}'_L}{M^u}' {\hat u}_R
+ \overline{{\hat d}'_L} {M^d}' {\hat d}_R + \mbox{h.c.}\Big{)}~,
\end{eqnarray}
where $\D\equiv \gamma^\mu D_\mu$ is the SM covariant derivative,
 $\phi$ is the Higgs doublet and $\phi^c$ is the charged conjugate
 one. 
In Eq.~(\ref{eq:lagrangian}), $Y$ and  $Y'$ are, respectively,
the square $n\times n$ and $m\times m$ Yukawa matrices; 
$M$ and  $M'$ are, respectively, the rectangular $n\times m$ and
$m\times n$ direct mass matrices.

Without loss of generality, the matrices $M$ and  $M'$ can always
be brought to the
$m\times m$ triangular form with  the rest being zero. Now,
one can rewrite the
Lagrangian~(\ref{eq:lagrangian}) in terms of the $m$  pairs
of the Dirac families $Q= (Q_L, Q'_R)$, constituting   the
vector-like representations of the SM,  and the $n-m$ chiral
families $Q_L$. In neglect of the Yukawa couplings, the Lagrangian of
the Dirac families  is explicitly $P$ invariant. Hence, of
those initial $n+m$ chiral families, the $2m$ ones  transform  after
mass diagonalization to $m$ pairs of the heavy
vector-like families (VLF's).\footnote{To be precise we call as VLF
the family mass eigenstate  which possesses the (approximate)
left-right symmetric SM interactions.} 
This is to be expected according to the survival
hypothesis~\cite{georgi} because the chirally conjugate families
lose their chiral protection. 
The unbalanced $n-m$  families  can
be considered  as the (approximate) pure chiral
ones. In practice, we suppose that the net number of the chiral
families is three and hence $n=3+m$. 
                                
We generalize the parameter counting for the
chiral families of Ref.~\cite{santa} to the case with extra VLF's.
It goes as is shown in Table 2. Here $G$ is the global symmetry of
the kinetic part of the Lagrangian~(\ref{eq:lagrangian}). It is
broken explicitly by the mass terms, only the residual symmetry $H =
U(1)$ of the baryon number  being left in
the  general case we consider.\footnote{The
degenerate cases leave more residual symmetries and  require
special consideration.} Hence, the transformations  of
$G/H$ can be used to absorb the spurious parameters in
Eq.~(\ref{eq:lagrangian})
leaving only the physical set ${\cal M}_{phys}$ of them. The last
four lines in Table~2 present the physical parameters for the
minimal SM and for the three its simplest extensions: the traditional
one with a normal family, the one with a mirror family  and  the
one with of a pair of the normal and mirror families.\footnote{The
first two cases are practically excluded by the SM self-consistency
requirements~\cite{pir}.} The last
case will be considered in detail in the next section.

\begin{table}[htbp]
\paragraph{Table 2}Parameter counting in the symmetry/electroweak
basis.
\vspace{1ex}

\begin{tabular}{|c|c|c|}
\hline 
Couplings&Moduli&Phases\\
and symmetries&&\\
\hline
$Y^u, Y^d, {Y^u}', {Y^d}',$&$2(n^2 + m^2)$&$2(n^2 + m^2)$\\ 
$M, {M^u}', {M^d}'$&$+ 3n m$&$+ 3n m$\\
\hline
$G = U(n)^3\times U(m)^3$&$-\frac{3}{2} [n(n-1) + m(m-1)]$&
$-\frac{3}{2} [n(n+1) + m(m+1)]$\\  
\hline
$H = U(1)$&$0$&$1$\\
\hline
${\cal M}_{phys}(n, m)$&$\frac{1}{2} (n+m)(n+m-1) $&
$\frac{1}{2} (n+m-2)(n+m-1)$\\
&$ + 2n m +2(n + m)$&$ +2n m$\\
\hline
${\cal M}_{phys}^{\mbox{\scriptsize SM}}(3, 0)$&$9 = 3 + 6$&$1$\\
\hline
${\cal M}_{phys}(4, 0)$&$ 8 + 6 = 14 $&$3$\\
\hline
${\cal M}_{phys}(3, 1)$&$ 8 + 12 = 20 $&$9$\\
\hline
${\cal M}_{phys}(4, 1)$&$10 + 18 = 28$&$14$\\
\hline
\end{tabular}
\end{table}

Further, the kinetic part of the effective Lagrangian with the $W$,
$Z$ and Higgs bosons
being integrated out is
\begin{eqnarray}
{\cal L}_{eff}&=&
~~i \overline{u_L}\D u_L + i \overline{d_L}\D d_L
+ i \overline{u_R}\D u_R
+ i \overline{d_R}\D d_R\nonumber\\
&&-\big(\overline{u_L}{\cal M}^{u}_{diag} u_R +   
\overline{d_L}{\cal M}^{d}_{diag} d_R +\mbox{h.c.} \big)~,  
\end{eqnarray}
where $\D$ means the covariant derivatives w.r.t.\ the QED and QCD
only; $ u_\chi$ and  $d_\chi$ ($\chi = L$, $R$) generically mean
the quarks  in the mass/flavour basis, and  ${\cal M}^{u, d}_{diag}$
are the diagonal mass matrices defining the basis.
The corresponding parameter counting is presented in Table~3. 
Due to the absence of mutual quark transitions,
the total residual symmetry of the mass matrices ${\cal
M}^{u,d}_{diag}$ is here $H = U(1)^{2(n+m)}$.
Table~3 clearly shows the breakdown  of the moduli of ${\cal
M}_{phys}$ in Table~2 on the physical masses and mixing angles.  

\begin{table}[htbp]
\paragraph{Table 3}Parameter counting for the effective Lagrangian.
\vspace{1ex}

\begin{tabular}{|c|c|c|}
\hline 
Couplings&Moduli&Phases\\
and symmetries&&\\
\hline
${\cal M}^{u}, {\cal M}^{d}$&$2(n + m)^2$&$2(n + m)^2$\\
\hline
$G = U(n + m)^4$&$-2 (n + m)(n + m - 1)$&$-2 (n + m)(n + m + 1)$\\
\hline
$H = U(1)^{2(n + m)}$&$0$&$2 (n + m)$\\
\hline
${\cal M}^{u}_{diag}, ~{\cal M}^{d}_{diag}$&$2 (n + m)$&$0$\\
\hline
\end{tabular}
\end{table}

Let us now redefine collectively quarks 
in the symmetry basis as
\def\k{\kappa}
${\hat \k}_\chi = {\hat u}_\chi$, ${\hat d}_\chi$
and these in  the mass basis, i.e.\ the quark eigenstates with
${\cal M}_{phys}$ being diagonal, as $\k_\chi = u_\chi$, $d_\chi$
($\chi =
L$, $R$). The bases are related by the unitary $(n+m)\times(n+m)$
transformations

\begin{equation}
{\hat\k}_\chi{}_{A} = 
({U^\k_\chi})^{ F}_{ A}
\,{\k_\chi}_{F}~,
\end{equation}
with the ensuing bi-unitary mass diagonalization
\begin{equation}
\label{4}
{U^{\k}_L}^\dagger {\cal M}^{k} U^{\k}_R = 
{\cal M}^{\k}_{diag} = \mbox{diag\,}(\overline{ m}^{\k}{}_f , 
\overline{ M}^{\k}{}_4, \dots ,\overline{ M}^{\k}{}_{n+m})~. 
\end{equation}
In the equations above, the indices  $A = A_L,A_R$; $A_L = 1,\dots,
n$; $A_R = n+1,\dots, n+m$ are those in the symmetry basis, and 
$F = f,4, \dots ,n+m$; $f = 1,2,3$ are indices in the mass basis.
It is assumed that $\overline{m}^\k{}_f \ll \overline{M}^\k{}_4,
\dots , \overline{ M}^\k{}_{n+m}$.

The matrices $U^\k_\chi$ satisfy the unitarity relations 
\begin{equation}
\label{6}
{U^{\k}_\chi}\,U^{\k}_\chi{}^{\dagger} =I
\end{equation} 
and
\begin{equation}\label{5}
{U^{\k}_\chi}^{\dagger} I_L {U^{\k}_\chi} 
+ {U^{\k}_\chi}^{\dagger}I_R  {U^{\k}_\chi} = I~,
\end{equation}
were $I_L$, $I_R$ are the projectors onto the normal and mirror
subspaces in the symmetry basis:
\begin{eqnarray}
I_L&=&\mbox{diag}\,
(\,\underbrace{1,\dots,1}_{n}\,;\underbrace{0,\dots,0}_{m}\,)~,
\nonumber\\
I_R&=&\mbox{diag}\,
(\,\underbrace{0,\dots,0}_{n}\,;\underbrace{1,\dots,1}_{m}\,)
\end{eqnarray}
with $I_L+I_R=I$ and $I_\chi^2=I_\chi$.
Let us also introduce their transformation to the mass basis
\begin{equation}
\label{eq:projector}
X^\k_\chi = U^\k_\chi{}^{\dagger} I_\chi U^\k_\chi~.
\end{equation}
($\k = u$, $d$ and  $\chi = L$, $R$). Clearly, $X^\k_\chi$ are
Hermitian
and satisfy the projector condition: 
$X^\k_\chi{}^2 = X^\k_\chi{}$ (but
note that $X_L^\k + X_R^\k\neq I$ in the notations adopted). 

Now, the charged current Lagrangian is
\begin{equation}\label{eq:L_W}
- {\cal L}_W = \frac{g}{\surd\overline{2}} W^+_\mu 
\sum_\chi \overline{u_\chi} \gamma^\mu V_\chi d_\chi + \mbox{h.c.}
\end{equation}
and the neutral current one is
\begin{equation}\label{eq:L_Z}
- {\cal L}_Z = \frac{g}{c} Z_\mu
\sum_{\k,\chi} \overline{\k_\chi} \gamma^\mu N^{\k}_\chi\, \k_\chi~,
\end{equation}
where $c\equiv\cos\theta_W$, with $\theta_W$ being the 
Weinberg mixing angle. 
The corresponding quark mixing matrices  for the charged
currents are
\begin{equation}\label{V_chi}
{V_\chi} =
{U^{u}_\chi}^{\dagger}I_\chi U^{d}_\chi~,
\end{equation}
and for the neutral currents with the operator $T_3-s^2 Q$
\begin{equation}\label{eq:N_chi}
N^\k_\chi=T^\k_3 X^\k_\chi-s^2 Q^\k_\chi~.
\end{equation}
Here one has for the electroweak isospin: $T_3^\k = 1/2$ at $\k=u$
and $-1/2$ at $\k=d$; for the electric charge:  $Q_{L,R}^\k\equiv
Q^\k I$ with $Q^\k =2/3$ at $\k=u$ and $-1/3$ at $\k=d$;
$s\equiv\sin\theta_W$.

The charged current mixing matrices 
$V_L$ and $V_R$ play the role of the generalized
CKM matrices. But contrary to the minimal SM case, they as well as
the neutral current mixing matrices $N^{\k}_\chi$ are non-unitary.
Namely, one gets by the unitarity relations~(\ref{6})
\begin{eqnarray}\label{14}
V_\chi V_\chi^\dagger &=&  X^u_\chi ~,\nonumber\\
V_\chi^\dagger V_\chi &=& X^d_\chi~, 
\end{eqnarray}
where $X_\chi^\k$ ($X_\chi^\k\neq I$ in general) are given by
Eq.~(\ref{eq:projector}). 
From  the considerations above, the representations for the
$V_\chi$ follow
\begin{equation}\label{eq:representation}
V_\chi=X^u_\chi S_\chi=S_\chi X^d_\chi 
\end{equation}
with the unitary matrices $S_\chi={U^u_\chi}^\dagger U^d_\chi$ and
the positive definite Hermitian matrices $X^\k_\chi$, only one in a
pair with fixed $\chi$ being independent, say, $X_\chi^d\equiv
S_\chi^\dagger X_\chi^u S_\chi$.
The decomposition (\ref{eq:representation}) is known to be unique. 
In a case where there are only the normal
families,  one gets $X_L^\k=I$ and $X_R^\k=0$, so that
$V_L$ is unitary, $V_L=S_L$, and $V_R=0$.  

It is seen that the neutral current matrices  $N^\k_\chi$ are not
independent of the charged current ones $V_\chi$. In fact, one can
convince oneself  that $V_\chi$ and the diagonal mass matrices
${\cal M}^{\k}_{diag}$ suffice to parametrize all the fermion
interactions in a general class of the SM extensions by means 
of the arbitrary numbers 
of the vector-like isodoublets and
isosinglets~\cite{lavoura}.
Indeed, in the case at hand, using the unitarity relations (\ref{5}),
one gets for the Yukawa Lagrangian in the unitary gauge
\begin{eqnarray}\label{L_Y}
- {\cal L}_Y &=&
\frac{H}{v} \sum_\k \overline{\k_L} \Big{(} X^\k_L {\cal M}^\k_{diag}
                    - 2 X^\k_L {\cal M}^\k_{diag} X^\k_R  
                    + {\cal M}^{\k}_{diag} X^\k_R \Big{)} \k_R
\nonumber\\
&& + \sum_\k \overline{\k_L} {\cal M}^{\k}_{diag} \k_R +
\mbox{h.c.}~,  
\end{eqnarray}
$H$ being the physical Higgs boson. 
It follows from the above expression
and Eqs.~(\ref{eq:L_Z}), (\ref{eq:N_chi})
that all the flavour changing neutral currents are induced entirely 
by the lack of unitarity of the charged current mixing matrices
$V_\chi$. In the case with only the normal
families  ($X^\k_L=I$, $X^\k_R=0$) the usual SM  expressions for
${\cal L}_W$, ${\cal L}_Z$ and ${\cal L}_Y$ are recovered, the two
latter ones being flavour conserving.

We propose the following prescription for the model independent
paramet\-ri\-za\-tion of the $V_\chi$. 
The problem is that they are non-unitary and thus are difficult 
to parametrize directly. So, the idea is to express them in 
terms of a set of the auxiliary unitary matrices.
First of all, note that in the absence of any restrictions on the 
Lagrangian the unitary matrices
$U^\kappa_\chi$ in Eq.~(3) would be arbitrary. Now, an arbitrary
$(n + m)\times (n + m)$ unitary matrix $U$ can always be uniquely 
decomposed as $U = U{\vert}_{n\times n} ~U{\vert}_{m\times m}
~U{\vert}_{n\times m}$. 
Here $U{\vert}_{n\times n}$ is a unitary matrix in the $n\times n$ 
subspace. It is built of the $n^2$ generators. Similarly, 
$U{\vert}_{m\times m}$ is the restriction of $U$ onto the $m\times m$
subspace, and it is built of the $m^2$ generators. And finally,
$U{\vert}_{n\times m}$ means a unitary $(n + m)\times (n + m)$
matrix built of the $2nm$ generators which mix the two subspaces.

Now, by means of the symmetry basis transformations $G$ of 
Table~2 one can always put, without loss of generality, the matrices
$U^\kappa_\chi$ to the form
\begin{eqnarray}\label{eq:U_repr}
U^u_L &=& U^u_L{\vert}_{n\times m}~,
\nonumber\\
U^u_R &=& U^u_R{\vert}_{n\times m}~,
\nonumber\\
U^d_L &=& U^d_L{\vert}_{n\times n} ~U^d_L{\vert}_{n\times m}~,
\nonumber\\
U^d_R &=& U^d_R{\vert}_{m\times m} ~U^d_R{\vert}_{n\times m}~.
\end{eqnarray}
This representation includes six auxiliary unitary matrices.
Clearly, they depend on the $[n(n-1)/2 + m(m-1)/2 + 4mn]$ moduli and
$[n(n+1)/2 + m(m+1)/2 + 4mn]$ phases, and these numbers are
redundant.
But the $nm$ moduli and the same number of phases can be eliminated 
through the $n\times m$ matrix constraint
\begin{equation}\label{eq:constraint}
I_L U^u_L {\cal M}^u_{diag} {U^u_R}^{\dagger} I_R =
I_L U^d_L {\cal M}^d_{diag} {U^d_R}^{\dagger} I_R~.
\end{equation}
The latter one follows from the equality of the direct mass matrices
$M$ in
Eq.~(1) for the up and down quarks, 
and it includes additionally the $2(n+m)$ independent
moduli which enter ${\cal M}^u_{diag}$ and ${\cal M}^d_{diag}$.
By means of  Eq.~(\ref{eq:constraint}) one can express, 
e.g., one of the 
$U^\kappa_\chi{\vert}_{n\times m}$ in terms of all other matrices.
And finally, the $2(n+m)-1$ phases can be removed via the residual
phase
redefinition for the quarks in the mass basis. Putting all together,
one can easily verify that the total number of the independent
parameters is precisely as expected from  Table 2.

Having parametrized the auxiliary unitary matrices, 
one gets for the $V_\chi$
\begin{eqnarray}\label{eq:V_repr}
V_L &=& {U^u_L}^\dagger{\vert}_{n\times m}~ I_L~
        {U^d_L}{\vert}_{n\times n} ~{U^d_L}{\vert}_{n\times m}~,
\nonumber\\
V_R &=& {U^u_R}^\dagger{\vert}_{n\times m}~ I_R~
        {U^d_R}{\vert}_{m\times m} ~{U^d_R}{\vert}_{n\times m}
\end{eqnarray}
and for the $X^\kappa_\chi$
\begin{equation}\label{eq:X_repr}
X^\kappa_\chi = {U^\kappa_\chi}^\dagger{\vert}_{n\times m}~ I_\chi~ 
                {U^\kappa_\chi}{\vert}_{n\times m}~.
\end{equation}
When eliminating the $2(n+m)-1$ redundant phases one can always
take such a choice as to render the diagonal and above-the-diagonal
elements of the $V_L$ (or $V_R$) to be real and positive.

This gives a principal solution to the problem. When there are only 
the normal families ($m = 0$) the usual parametrization in terms of
just one unitary matrix $U^d_L{\vert}_{n\times n}$ is readily
recovered.
For the case with a pair of VLF's ($n=4$, $m=1$) we got also the
explicit expressions
of all the relevant quantities in terms of a minimal common set of
the independent arguments parametrizing the mass matrices (see the
next section). It is of much use at the model independent
parametrization to estimate the relative magnitudes of the various
mixing elements in terms of a small quantity 
$\epsilon = v^2/M^2\ll 1$. Otherwise, one has a priori no idea of
this. 

Finally, under small mixing it is useful to  decompose
\begin{equation}
V_\chi = V_{0 \chi} + \Delta V_{\chi}~,
\end{equation}
with the decoupling limit taken as  the zeroth order approximation
$V_{0 \chi}$, and with corrections 
$\Delta  V_{\chi}$  vanishing at $M\gg v$. 
To illustrate the behavior in the   limit, let us consider
the aforementioned case with a pair of the VLF's. One gets here 
\begin{equation}\label{eq:V_0L}
V_{0L}= \left({
\begin{array}{ccc}
V_{C}&0&0\\
0&1&0\\
0&0&0
\end{array}
}\right)
\end{equation}
and 
\begin{equation}\label{eq:V_0R}
V_{0R} = \mbox{diag\,}(0,0,0,1,0)~,
\end{equation}
$V_{C}$ being the usual  $3\times 3$ charged current 
matrix of the SM. Hence, for the $X^\k_\chi$ as given by
Eq.~(\ref{14})
one has in the zeroth order
\begin{eqnarray}\label{eq:declim}
X^{u,d}_{0L}&=&\mbox{diag\,}(1,1,1,1,0)~,\nonumber\\
X^{u,d}_{0R}&=&\mbox{diag\,}(0,0,0,1,0)~.
\end{eqnarray}
It follows from Eqs.~(\ref{eq:V_0L})--(\ref{eq:declim}) that in the
limit $M\gg v$ there are indeed two VLF's, the forth and the fifth
ones,  that interact in the left-right symmetric manner,  one of the
VLF's, the fifth one, being singlet under interactions with the $W$
boson. Besides, as it follows from Eq.~(\ref{L_Y}), both these
families decouple from the Higgs boson in the leading 
order of ${\cal O}(M/v)$, only the Yukawa terms ${\cal O}(M^0)$ being
left at most.

\section{Explicit realization}

The mass/flavour basis parameters, ${\cal M}^{u,d}_{diag}$ and
$V_{L,R}$, are phenomenological by their very nature.
They reflect an obscure mixture of contributions of  quite a
different physical origin. In particular, they shed no light on
the mixing magnitudes. On the contrary, the parameters in the
symmetry basis,
i.e.\ Yukawa couplings and the direct mass terms $M$ and
${M^u}'$, ${M^d}'$,
have the straightforward theoretical meaning. So, we express the
former ones in terms of the latter ones. This permits us to expand
upon the idea of the relative magnitude of the various mixing
elements in terms of the small quantity $v/M$.

The asymptotic freedom requirement for the $\mathrm SU(2)_W$
electroweak interactions results in the restriction that the total
number of the electroweak doublets 
should not exceed 21. The number of doublets in a chiral family being
4,  this is equivalent to the restriction that the total number of 
the families is $(n+m)\le 5$. 
Hence the maximum number of the extra VLF's allowed by the asymptotic
freedom is two, the case we stick to in what follows.\footnote{This 
might be a landmark for the number of the extra families.}

Using here the global symmetries $G$ of the Table~2 one can bring,
without loss of generality,  the
quark mass matrices in the symmetry basis to the following   
canonical form 
\begin{equation}\label{20}
{\cal M}^\k = \left(
\begin{array}{ccc}
\vspace{0.5ex}
{m^\k}^g_f&{\mu^\k}'_f&0\\
\vspace{0.5ex}
{\mu^\k}^g&m^\k{}_4&M\\
0&{M^\k}'&m^\k{}_5
\end{array}
\right)~,
\end{equation}
where $M$, ${M^\k}'$ are the real scalars and $\mu^\k{}^f$,
${\mu^\k}'_f$,
$m^\k{}_4$, $m^\k{}_5$ are in general complex. 
Here the lower case characters generically mean the masses of 
the Yukawa origin ($\sim Yv$).
Let us remind that $M$ in Eq.~(\ref{20}) is common for both
${\cal M}^u$ and ${\cal M}^d$.
The three-dimensional matrices $m^\k$ are Hermitian and positive
definite,  and one of them, e.g. $m^u$, can always be  chosen
diagonal. Under such a choice one can  simplify further:
\begin{equation}\label{21}
{\cal M}^\k_0 = {U^\k_0}^\dagger {\cal M}^\k U^\k_0, 
\end{equation}
where
\begin{equation}\label{eq:M_0_kappa}
{\cal M}_0^{\k} = \left(\begin{array}{ccccc}
\vspace{0.5ex}
m^\k{}_1&0&0&{\mu^\k}'_1&0\\
\vspace{0.8ex}
0&m^\k{}_2&0&{\mu^\k}'_2&0\\
\vspace{0.5ex}
0&0&m^\k{}_3&{\mu^\k}'_3&0\\
\vspace{0.5ex}
{\mu^\k}^1&{\mu^\k}^2&{\mu^\k}^3&m^\k{}_4&M\\
\vspace{0.5ex}
0&0&0&{M^\k}'&m^\k{}_5     
\end{array}\right)
\end{equation}
\noindent
with a redefinition of $\mu^\k{}^f$ and ${\mu^\k}'_f$, and with the
diagonal elements 
$m^\k{}_f$ being real and positive. The  matrices ${\cal M}_0^{\k}$
have a lot of texture zeros and are easiest to operate. 
The corresponding unitary $U^\k_0$ are
given by 
\begin{eqnarray}\label{22a}
U^{u}_0 &=& I~,\nonumber\\
U^d_{0 } &=& \left({
\begin{array}{cc}
V_{C}&0\\
0&I_2
\end{array}
}\right)~,
\end{eqnarray}
$V_C$ being the $3\times 3$ CKM matrix  and $I_2$ the $2\times 2$
identity matrix.  The mass matrices of Eq.~(\ref{eq:M_0_kappa})
possess the residual symmetry $U(1)^6$ which is reduced
to  $U(1)^5$  by the baryon number conservation.
So, one can use the phase redefinitions for two of the light
$d$ quarks  which leave just one complex phase in $V_C$ in accordance
with the decoupling limit requirement. 

It is seen from Eqs.~(\ref{eq:M_0_kappa}) and (\ref{22a}) that in
this parametrization the total number of  physical moduli is 
$10 + 15 + 3 = 28$ as it should be according to Table~2.
As for the phases, their number is in general $16 + 1 = 17$,
i.e.\ three of them are spurious
and can be removed. For example, by means of the residual phase
redefinition for the  three
light $u$ quarks one can make $\mu^u{}^f$ or ${\mu^u}'_f$ to be real,
or put some other three relations on their phases.
This exhausts the freedom of the phase redefinitions, leaving only
the physical parameters. 

The characteristic equations (see Appendix) 
\begin{equation}\label{23}
\det\,({\cal M}^\k_0 {{\cal M}^\k_0}^\dagger - \lambda^\k I) = 0
\end{equation}
give for the roots in the first order 
(i.e. up to the relative
corrections ${\cal O}(v^2/M^2)$ to the leading order):
\begin{eqnarray}\label{eq:lambda}
\lambda_f\equiv \overline{ m}_f^2 &=&m_f^2\bigg{(}1
 - \Big(\frac{\vert\mu^f\vert^2}{M^2} +
\frac{\vert\mu'_f\vert^2}{M^{'2}}\Big)\bigg{)}
 + \frac{m_f}{M M'} (m_5\mu^f\mu'_f + \mbox{h.c.})~,\nonumber\\
\lambda_4\equiv \overline{ M}_4^2&=&M^2 
+ \Sigma \vert{\mu^f}\vert^2+ \vert m_4\vert^2 +
\vert{m_5}\vert^2\nonumber\\
&&+ \frac{M'{}^2}{M^2 - M'{}^{2}} \Big{(}  (\vert{m_4}\vert^2 +
\vert{m_5}\vert^2)
   + \frac{M}{M'}(m_4 m_5 + \mbox{h.c.})\Big{)}~,\nonumber\\
\lambda_5\equiv \overline{ M}_5^2&=&M^{'2}
+ \Sigma\vert\mu'_f\vert^2 + \vert m_4\vert^2 + \vert
m_5\vert^2\nonumber\\
&&+ \frac{M^2}{M'{}^{2} - M^2} \Big{(} (\vert m_4\vert^2 + \vert
m_5\vert^2)
   + \frac{M'}{M} (m_4 m_5 + \mbox{h.c.})\Big{)}
\end{eqnarray}
with the superscripts $\k = u$, $d$ being suppressed.\footnote{Hence,
the up and down quarks of the fourth family are always (almost)
degenerate, whereas those of the fifth family are in general not.
Nevertheless, because the fifth family does not couple to the $W$
boson in the zeroth order (see Eqs.~(21), (22)) this does not result
in the strong coupling $\sim ({M^u}' - {M^d}')$ of the longitudinal
$W$ with the fifth heavy family, as well as with the fourth one.
}
Here it is supposed that one has, in general,  $M\sim M'$ but 
$M\neq M'$.\footnote{The degenerate case $M = {M^\kappa}'$ (for one
or both $\kappa = u,d$) is to be studied separately.
It modifies the results for heavy families, but fortunately does not
influence the validity of those concerning the light quarks
exclusively.}

It is seen that corrections to $m_f^2$ are proportional to $m_f$ 
themselves, i.e. the light quarks are  still chirally protected.
This property drastically reduces the otherwise dangerous corrections
to the masses of the lightest $u$ and $d$ quarks at the moderate $M$.
In the limit $m_f\to 0$
it naturally happens without any fine tuning beyond that of the SM.
On the other hand, it means that within the perturbation theory
the masses of the lightest quarks cannot
entirely be induced by an admixture of the vector-like families:
if $m_f = 0$ then $\overline{m}_f = 0$, too.
But at the finite $m_f$ one finds for the masses of the light quarks
\begin{equation}\label{eq:m_light}
\overline{m}_f = m_f \Bigg{(} 1 
- \frac{1}{2} \Big{(} \frac{\vert\mu^f\vert^2}{M^2} 
+ \frac{\vert\mu'_f\vert^2}{{M'}^2}
\Big{)}\Bigg{)} 
+ \frac{1}{2} \Big{(} \frac{m_5\mu^f \mu'_f}{MM'} + \mbox{h.c.}
\Big{)}~,
\end{equation}
and for the validity of perturbative expansion it could require some
fine tuning for $m_5$ at the moderate $M$. 
   
Once the physical 
masses are known, one can obtain the matrices $U^\k_{1L}$ and 
$U^\k_{1R}$ of the bi-unitary transformation 
\begin{equation}\label{26}
U^{\k\dagger}_{1L} {\cal M}^\k_0 U^\k_{1R} = {\cal M}^\k_{diag}~.
\end{equation}
Obviously, they satisfy the relations
\begin{eqnarray}\label{27}
{{\cal M}^\k_0}^\dagger {\cal M}^\k_0 U^\k_{1R} &=&
U^\k_{1R} {{\cal M}^\k_{diag}}^2~,\nonumber\\
{\cal M}^\k_0 {{\cal M}^\k_0}^\dagger U^\k_{1L} &=&
U^\k_{1L} {{\cal M}^\k_{diag}}^2
\end{eqnarray}

\noindent
which are to be considered as the sets of the independent linear
equations
for their columns. Having solved the equations, one can find the
elements of $U^\k_{1\chi}$ which are given in  Appendix.
Finally, one has for the total matrices of the bi-unitary
transformations of Eq.~(\ref{4})
\begin{equation}\label{27a}
U^\k_\chi = {U^\k_0}\, U^\k_{1\chi}~,
\end{equation}
where $U^\k_0$ are given by Eq.~(\ref{22a}).

Hereof one gets the mixing matrices $V_\chi$  given by 
Eqs.~(\ref{A9}), (\ref{A10}) of Appendix and then the charged
current Lagrangian ${\cal L}_W$ given by Eq.~(\ref{eq:L_W}).
The $Z$-mediated neutral current Lagrangian ${\cal L}_Z$ is  given
by Eqs.~(\ref{eq:L_Z}), (\ref{eq:N_chi}) with $X^\kappa_\chi$
from Eqs.~(\ref{A11}), (\ref{A12}).
The neutral scalar current Lagrangian takes the general form
\begin{equation}\label{29}
- {\cal L}_H = \frac{H}{v} \sum_\k 
\overline{\k_L}\, {U^\k_L}^\dagger ({\cal M}^\k 
- {\cal M}^\k_{dir}) U^\k_R \, \k_R + \mbox{h.c.}
\end{equation}
with the direct mass matrices
\begin{equation}\label{30}
{\cal M}^\k_{dir} = \left({
\begin{array}{ccc}
O_3&0&0\\
0&0&M\\
0&{M^\k}'&0
\end{array}
}\right)~,
\end{equation}
where $O_3$ is the $3\times 3$ zero matrix. As a consequence
of the
substraction of the direct mass terms, the total mass and Yukawa
matrices are not diagonalizable simultaneously in the same basis, 
at variance with the SM case.
In the mass basis,  the Higgs interaction Lagrangian is non-diagonal
\begin{eqnarray}\label{eq:Higgs_Lagr}
- {\cal L}_H &=& \frac{H}{v} \sum_\k {\overline{\k_L}}~ {\cal H}^\k
\k_R + \mbox{h.c.}
\end{eqnarray}
with the explicit form of ${\cal H}^\kappa$ given 
by Eqs.~(\ref{A13}), (\ref{A14}) of Appendix.

One should stress  that for the light quarks all the off-diagonal 
components of the  Lagrangian  ${\cal L}_W$ (beyond that of the
minimal SM), as well as those of the
${\cal L}_Z$ and ${\cal L}_H$ are suppressed by the ratio $v^2/M^2$,
and it does not depend on the details of the mass matrices.
Besides, it follows from the above that, among the off-diagonal
interactions, the Higgs mediated interactions are the only ones that
do not vanish in the decoupling limit.
Hence, the heavy quarks are expected to decay mainly into the light
ones and the Higgs boson with the natural decay width 
$\Gamma\sim \vert{}Y\vert^2/4\pi~M$. 
As a result, all the leading loop corrections to the light quark 
processes with the internal heavy vector-like quarks are expected 
to be mediated by the Higgs boson exchanges. 
So, the modern SM physics, i.e. predominantly that of the 
light fermions and the gauge bosons, may be succeeded by that of 
the heavy vector-like fermions and the Higgs boson.

\section{Conclusions}

We have shown that the mere addition of a pair of the VLF's
drastically changes all the characteristic features of the minimal
SM. First of all, the generalized CKM matrix for the  left-handed
charged currents ceases
to be unitary. Moreover, this non-unitarity takes place in the whole 
flavour space but not only in the light quark sector which would
occur for adding  only the normal families.
Further, there appear the right-handed charged currents,
the flavour changing neutral currents, both the vector and  
scalar ones, all with the non-unitary mixing matrices and with a
number of $CP$ violating phases. 

Due to decoupling relative to  the large direct mass terms $M$, the
extended SM definitely does not contradict  experiment in the 
limit $M\gg v$. But at the moderate  $M>v$, the addition of a pair of
the VLF's would make the model phenomenology, especially that 
of the flavour and $CP$ violation, extremely diverse. 
So, the extension opens new prospects for studying the deviations
from the SM in the future experiments at high energies.

\section*{ Appendix }

\setcounter{equation}{0}
\def\theequation{A.\arabic{equation}}

One has generically (with the indices $\k = u$, $d$ being omitted)
\begin{equation}\label{A1}
\begin{array}{l}
{\cal M}_0 {\cal M}_0^\dagger = \\
\\
\left({
\begin{array}{ccccc}
\vspace{1ex}
(m_1^2+\vert\mu'_1\vert^2)&\mu'_1 {\mu'_2}^*&\mu'_1 {\mu'_3}^*&
(m_1\mu_1^*+\mu'_1 m_4^*)&\mu'_1 {M'}\\
\vspace{1ex}
\mu'_2 {\mu'_1}^*&(m_2^2+\vert\mu'_2\vert^2)&\mu'_2 {\mu'_3}^*&
(m_2\mu_2^*+\mu'_2 m_4^*)&\mu'_2 {M'}\\
\vspace{1ex}
\mu'_3 {\mu'_1}^*&\mu'_3 {\mu'_2}^*&(m_3^2+\vert\mu'_3\vert^2)&
(m_3\mu_3^* + \mu'_3 m_4^*)&\mu'_3 {M'}\\
\vspace{1ex}
(m_1\mu_1\phantom{+}&(m_2\mu_2\phantom{+}&(m_3\mu_3\phantom{+}&({
M}^2+{\vert
m_4\vert}^2&(m_4 {M'}\phantom{+}\\
\vspace{1ex}
~+{\mu'_1}^* m_4)&+{\mu'_2}^* m_4)&+{\mu'_3}^* m_4)&
+\Sigma{\vert\mu^f\vert}^2)& + M m_5^*)\\
\vspace{1ex}
{\mu'_1}^* M'&{\mu'_2}^* M'&{\mu'_3}^* M'&(m_4^* M'+M m_5)&({
M'}^2 + {\vert m_5\vert}^2) \\
\end{array}
}\right)\,.
\end{array}
\end{equation}

\noindent
The characteristic equation
\begin{equation}\label{A2}
\det\,({\cal M}_0 {\cal M}_0^{\dagger} - \lambda I) = 0
\end{equation}
in the explicit form is
\begin{eqnarray}\label{A3}
&&  \lambda^5 - \lambda^4 \bigg[M^2 + {M'}^2 
              + \Sigma \Big{(} m_f^2 + \vert\mu^f\vert^2 
              + \vert\mu_f'\vert^2 \Big{)} 
              + \vert m_4\vert^2 
              + \vert m_5\vert^2\bigg] 
\nonumber\\&& + \lambda^3 \bigg{[}
                  M^2 {M'}^2 + M^2 \Sigma \Big{(} m_f^2 
                            + \vert\mu_f'\vert^2\Big{)} 
               + {M'}^2 \Sigma \Big{(} m_f^2 
                            + \vert\mu^f\vert^2\Big{)}
\nonumber\\&& - M {M'} (m_4 m_5 + \mbox{h.c.})
              + m_1^2 m_2^2 + m_1^2 m_3^2 + m_2^2 m_3^2\bigg{]}
\nonumber\\&& - \lambda^2 \bigg{[}M^2 {M'}^2 \Sigma m_f^2 
                 + M^2 \Big{(}m_1^2 m_2^2 + m_1^2 m_3^2 + m_2^2 m_3^2  
\nonumber\\&&    + m_1^2 (\vert\mu_2'\vert^2 + \vert\mu_3'\vert^2) 
                 + m_2^2 (\vert\mu_1'\vert^2 + \vert\mu_3'\vert^2) 
                 + m_3^2 (\vert\mu_1'\vert^2 + \vert\mu_2'\vert^2)
                 \Big{)}
\nonumber\\&& + {M'}^2 \Big{(}m_1^2 m_2^2 + m_1^2 m_3^2 + m_2^2 m_3^2 
\nonumber\\&& + m_1^2 (\vert{\mu}_2\vert^2 + \vert{\mu}_3\vert^2) 
                 + m_2^2 (\vert{\mu}_1\vert^2 + \vert{\mu}_3\vert^2) 
                 + m_3^2 (\vert{\mu}_1\vert^2 + \vert{\mu}_2\vert^2)
                 \Big{)}
\nonumber\\&& + M {M'} \Big{(}( - m_4 \Sigma m_f^2 
                 + \Sigma m_f\mu^f\mu'_f) m_5 + \mbox{h.c.} \Big{)} 
                 + m_1^2 m_2^2 m_3^2 \bigg{]}
\nonumber\\&& + \lambda \bigg{[} M^2 {M'}^2 \Big{(} m_1^2 m_2^2 +
m_1^2 m_3^2 
                          + m_2^2 m_3^2 \Big{)}
\nonumber\\&& + M^2 \Big{(} m_1^2 m_2^2 m_3^2 
                     + m_2^2 m_3^2 \vert\mu_1'\vert^2 
                     + m_1^2 m_3^2 \vert\mu_2'\vert^2 
                     + m_1^2 m_2^2 \vert\mu_3'\vert^2 \Big{)}
\nonumber\\&& + {M'}^2 \Big{(} m_1^2 m_2^2 m_3^2 
              + m_2^2 m_3^2 \vert\mu_1\vert^2 
              + m_1^2 m_3^2 \vert\mu_2\vert^2 
              + m_1^2 m_2^2 \vert\mu_3\vert^2 \Big{)} \bigg{]}
\nonumber\\&& - \bigg{[}M^2 {M'}^2 m_1^2 m_2^2 m_3^2
              + M {M'} \Big{(}\big{(} 
               - m_1^2 m_2^2 m_3^2 m_4  
               + m_1 m_2^2 m_3^2 \mu_1\mu'_1  
\nonumber\\&&  + m_1^2 m_2 m_3^2 \mu_2\mu'_2 
               + m_1^2 m_2^2 m_3 \mu_3\mu'_3 \big{)} m_5  
               + \mbox{h.c.} \Big{)}\bigg{]} + \dots = 0~.
\end{eqnarray}

\noindent
Let us rewrite it in terms of the dimensionless quantity
$x\equiv\lambda/M^2$. Then, one can transform Eq.~(\ref{A3}) as
\begin{equation}\label{A4}
\Big{[}\prod_f (x - x^{(0)}_f)\Big{]} (x - x^{(0)}_4)(x - x^{(0)}_5)
= \epsilon P_4(x)~,
\end{equation}

\noindent
where
\begin{equation}\label{A5}
\epsilon = \frac{1}{M^2} \Big{(}\Sigma\vert\mu^f\vert^2 
+ \Sigma\vert\mu'_f\vert^2 + \vert m_4\vert^2 + \vert m_5\vert^2
\Big{)} 
\end{equation}
is the small paremeter ($\epsilon = {\cal O}(v^2/M^2)$)
and  $x^{(0)}_f\equiv m_f^2/M^2 = {\cal O}(\epsilon)$,
$x^{(0)}_4 = 1$,  $x^{(0)}_5 = {M'}^2/M^2$
are the zeroth order roots. The fourth
power polynomial
$P_4(x) = (x^4 + \dots)$ has coefficients ${\cal O}(1)$ or less.
The dropped out terms corresponding to dots in Eq.~(\ref{A3}) result
in
the relative corrections ${\cal O}(\epsilon^2)$, 
and hence they can be omitted in our approximation.
Iterating Eq.~(\ref{A4}) one arrives at the roots of
Eq.~(\ref{eq:lambda}).

The elements of the $U_{1L}$ matrix (with the indices $\k = u$, $d$
being suppressed) are as follows
\begin{eqnarray}\label{A6}
&&U_{1L}{}^f_g = \delta^f_g\bigg{(} 1 - \frac{1}{2 M^2}n^f_f\bigg{)} 
 + (\delta^f_g - 1) \frac{1}{M^2} p^f_g~,\nonumber\\
\vspace{0.5ex}
&&U_{1L}{}^f_4 = \frac{1}{M^2} p^f_4~~,~~
U_{1L}{}^f_5 = \frac{1}{M} p^f_5~,\nonumber\\
\vspace{0.5ex}
&&U_{1L}{}_f^4 = \frac{1}{M^2} p^4_f~~,~~
U_{1L}{}_f^5 = \frac{1}{M} p^5_f~,\nonumber\\
\vspace{0.5ex}
&&U_{1L}{}_5^4 = \frac{1}{M} p^4_5~~,~~
U_{1L}{}_4^5 = \frac{1}{M} p^5_4~,\nonumber\\
\vspace{0.5ex}
&&U_{1L}{}_4^4 = 1 - \frac{1}{2M^2} n_4^4~~,~~
U_{1L}{}_5^5 = 1 - \frac{1}{2M^2} n_5^5~,\nonumber\\
\end{eqnarray}
and
\begin{eqnarray}\label{A7}
&&U_{1R}{}^f_g =
\delta^f_g\bigg{(} 1 - \frac{1}{2{M'}^2} {n'}^f_f\bigg{)} 
 + (\delta^f_g - 1) \frac{1}{{M'}^2} {p'}^f_g~,\nonumber\\
\vspace{0.5ex}
&&U_{1R}{}^f_4 = \frac{1}{{M'}^2} {p'}^f_4~~,~~
U_{1R}{}^f_5 = \frac{1}{M'} {p'}^f_5~,\nonumber\\
\vspace{0.5ex}
&&U_{1R}{}_f^4 = \frac{1}{{M'}^2} {p'}^5_f~~,~~
U_{1R}{}_f^5 = \frac{1}{M'} {p'}^4_f~,\nonumber\\
\vspace{0.5ex}
&&U_{1R}{}_4^4 = \frac{1}{M'} {p'}^5_4~~,~~
U_{1R}{}_5^5 = \frac{1}{M'} {p'}^4_5~,\nonumber\\
&&U_{1R}{}_5^4 = 1 - \frac{1}{2{M'}^2} {n'}_5^5~~,~~
U_{1R}{}_4^5 = 1 - \frac{1}{2{M'}^2} {n'}_4^4~,
\end{eqnarray}
where
\begin{eqnarray}\label{A8}
p^f_g &=&                          
\frac{\mu^f (m_f^2-\vert m_5\vert^2)(m_f {\mu^f}^*\mu'_g -
   m_g {\mu^g}^*\mu'_f)
         + k_f (m_f\mu'_g - \frac{m_g}{m_f}\frac{M'}{M} {\mu^g}^*
         m_5^*)}
        {(m_g^2-m_f^2)(m_f\mu'_f - \frac{M'}{M}
        m_5^* {\mu^f}^*)}~,\nonumber\\ 
\vspace{0.5ex}\nonumber\\		
p^f_4 &=&
- k_f ~\frac{\Big(k_f+\vert\mu^f\vert^2 (m_f^2-\vert m_5\vert^2)
  \Big)\Big(\frac{M'}{M} m_5^*
  + \frac{1}{k_f} m_f\mu^f\mu'_f (m_f^2-\vert
  m_5\vert^2)\Big)} 
       {m_f(m_f\mu'_f - \frac{M'}{M} m_5^* {\mu^f}^*)}~,\nonumber\\
\vspace{0.5ex}\nonumber\\
p_f^4 &=&
m_f {\mu^f}^* - \frac{\mu'_f (\rho + \vert m_5\vert^2)} 
                  {m_4 + \frac{M'}{M} m_5^*}
~,\nonumber\\ 
\vspace{0.5ex}\nonumber\\
p^f_5 &=&
\frac{\frac{M'}{M} (k_f+m_f^2\vert\mu^f\vert^2) - m_f m_5\mu^f\mu'_f}
     {m_f(m_f\mu'_f - \frac{M'}{M} m_5^* {\mu^f}^*)}~~~,~~~
p_f^5 =
\frac{M}{M'} \mu'_f~,\nonumber\\
\vspace{0.5ex}\nonumber\\
p_5^4 &=&
\frac{m_4 m_5 - \frac{M'}{M}\rho}{m_4 + \frac{M'}{M} m_5^*}~~~,~~~
p_4^5 =
\frac{MM'}{{M'}^2-M^2} \Big{(} m_4 + \frac{M}{M'} m_5^*\Big{)}
~,\nonumber\\
\vspace{0.5ex}\nonumber\\
n^f_f &=&
\bigg{\vert} 
\frac{\frac{M'}{M} (k_f+m_f^2\vert\mu^f \vert^2) 
       - m_f m_5\mu^f\mu'_f}
     {m_f(m_f\mu'_f - \frac{M'}{M} m_5^* {\mu^f}^*)}
\bigg{\vert}^2~,\nonumber\\
\vspace{0.5ex}\nonumber\\
n^4_4 &=& 
\bigg{\vert}
\frac{m_4 m_5 - \frac{M'}{M} \rho}
     {m_4 + \frac{M'}{M} m_5^*}
\bigg{\vert}^2~,\nonumber\\
\vspace{0.5ex}\nonumber\\				   
n^5_5 &=& 
\Big{\vert} 
\frac{{M'}^2}{{M'}^2-M^2} (m_4 + \frac{M}{M'} m_5^*)
\Big{\vert}^2
+ \Sigma \vert\mu'_f\vert^2~
\end{eqnarray}
and $k_f = M^2 (\overline{m}_f^2 - m_f^2)$,
$\rho = M^2 + \Sigma{\vert\mu^f\vert}^2 - \overline{ M}_4^2$.
The $p'$, $n'$ are obtained from $p$, $n$, respectively, by 
substituting $\mu^f\leftrightarrow {\mu'_f}^*$, 
$m_4\leftrightarrow m_4^*$, $m_5\leftrightarrow m_5^*$,
$M\leftrightarrow M'$. 
All these auxiliary parameters are in general of order ${\cal
O}(M^0)$. 
The elements of the matrix $U_{1R}$ 
are obtained from those for $U_{1L}$ by the same substitution 
followed by changing column indices $4\leftrightarrow 5$ for 
the matrix elements  $(U_{1L})^4_A$ and $(U_{1L})^5_A$.

Hereof one gets for the charged current matrix
$V_L = {V_0}_L + \Delta V_L$ 
\def\u{{p^u}{}}
\def\d{{p^d}{}}
\begin{equation}\label{A9}
\begin{array}{l}
\Delta V_L = 
\\
\\
\left({
\begin{array}{c|cc}
- \frac{1}{M^2} \sum \Big{(} {\u^f_h}^* {V_C}^g_h + {V_C}_f^h \d^g_h
\Big{)}&
\frac{1}{M^2} \Big{(} \sum {V_C}_f^h \d^4_h + {\u^f_4}^* \Big{)}&
\frac{1}{M} \sum {V_C}_f^h \d^5_h \\
- \frac{1}{2M^2} ({n^u}{}^f_f + {n^d}{}^g_g) {V_C}_f^g&
&\\
\cline{1-3}
&&\\
\frac{1}{M^2} \Big{(} \sum {\u^4_h}^* {V_C}_h^g + \d^g_4 \Big{)}&
 - \frac{1}{2M^2} ({n^d}^4_4 + {n^u}^4_4)&
\frac{1}{M} \d^5_4 \\
&&\\
\frac{1}{M} \sum {\u^5_h}^* {V_C}_h^g&
\frac{1}{M} {\u^5_4}^*&
\frac{1}{M^2} \Big{(} \sum {\u^5_h}^* \d^5_k {V_C}^k_h\\
&
&
+ {\u^5_4}^* \d^5_4 \Big{)}
\end{array}
}\right)\,, 
\\
\\
\end{array}
\end{equation}
with ${V_0}_L$ from Eq.~(\ref{eq:V_0L}) 
and similarly for $V_R = {V_0}_R + \Delta V_R$ with 
${V_0}_R = \mbox{diag}~ (0,0,0,1,0)$ and
\def\u{{p^u}'{}}
\def\d{{p^d}'{}}
\begin{equation}\label{A10}
\begin{array}{l}
\Delta V_R = 
\left({
\begin{array}{c|cc}
\frac{1}{{M^u}' {M^d}'} {{\u}^f_5}^* {\d}^g_5&
\frac{1}{{M^u}'} {\u^f_5}^*&
\frac{1}{{M^u}' {M^d}'} {{\u}^f_5}^* {\d}^4_5\\
&&\\
\cline{1-3}
&&\\
\frac{1}{{M^d}'} {\d}^g_5&
 - \frac{1}{2{{M^u}'}^2} {{n'}^u}{}^5_5 - \frac{1}{2{{M^d}'}^2}
 {{n'}^d}{}^5_5&
\frac{1}{{M^d}'} {\d}^4_5 \\
&&\\
\frac{1}{{M^u}' {M^d}'} {{\u}^4_5}^* {\d}^g_5&
\frac{1}{{M^u}'} {{\u}^4_5}^*&
\frac{1}{{M^u}' {M^d}'} {{\u}^4_5}^* {\d}^4_5
\end{array}
}\right)~~.
\\
\end{array}
\end{equation}
For the neutral current matrices ($\kappa = u, d$ being suppressed
everywhere below) one gets
\def\p{p}
\begin{equation}\label{A11}
\begin{array}{ll}
X_L = X_{0L} - 
\left({
\begin{array}{ccc}
\frac{1}{M^2} {\p^f_5}^* \p^g_5&
\frac{1}{M^2} {\p^f_5}^* \p^4_5&
\frac{1}{M} {\p^f_5}^* \\
&&\\
\frac{1}{M^2} {\p^4_5}^* \p^g_5&
\frac{1}{M^2} \vert\p^4_5\vert^2&
\frac{1}{M} {\p^4_5}^*\\
&&\\
\frac{1}{M} \p^g_5&
\frac{1}{M} \p^4_5&
 - \frac{1}{M^2} n^5_5
\end{array}
}\right)~~,
\\
\end{array}
\end{equation}
and
\def\pk{{p'}{}}
\begin{equation}\label{A12}
\begin{array}{l}
X_R = X_{0R} +
\left({
\begin{array}{ccc}
\frac{1}{{M'}^2} {\pk^f_5}^* \pk^g_5&
\frac{1}{M'} {\pk^f_5}^*&
\frac{1}{{M'}^2} {\pk^f_5}^* \pk^4_5\\
&&\\
\frac{1}{M'} \pk^g_5&
 - \frac{1}{{M'}^2} {n'}^5_5&
\frac{1}{M'} \pk^4_5 \\
&&\\
\frac{1}{{M'}^2} {\pk^4_5}^* \pk^g_5&
\frac{1}{M'} {\pk^4_5}^*&
\frac{1}{{M'}^2} \vert\pk^4_5\vert^2
\end{array}
}\right)~~,
\\
\end{array}
\end{equation}
with $X_{0L} = I_L$ and $X_{0R} = \mbox{diag}~ (0,0,0,1,0)$.

Finally, for the Higgs mediated neutral current matrix
${\cal H} = {\cal H}_0 + \Delta {\cal H}$ one has
\def\k{{}} 
\begin{equation}\label{A13}
{\cal H}_0 = 
~ \left(
{
\begin{array}{ccc}
\overline{m}^\k_f\delta^g_f& 
0&
- \frac{M'}{M} {p^f_5}^*\\
&&\\
- \frac{M}{M'} {p'}^g_5&
0& 
- \frac{M'}{M} {p^4_5}^* - \frac{M}{M'} {p'}^4_5\\
&&\\
0&
- ({p^5_4}^* + {p'}^5_4)&
0\\
\end{array}
}\right)
\end{equation}
and
\begin{equation}\label{A14}
\begin{array}{l}
\Delta {\cal H} = 
\\ 
\\
\left(
{
\begin{array}{c|cc}
- \frac{1}{MM'} \bigg{(} {p^f_4}^* {p'}^g_5 + {p^f_5}^*
{p'}^g_4\bigg{)}&
- \frac{1}{M} \bigg{(} {p^f_5}^* {p'}^5_4 + {p^f_4}^*\bigg{)}&
\frac{1}{MM'} \bigg{(} \frac{1}{2} {p^f_5}^* {{n'}^\k}^4_4
- {p^f_4}^* {p'}^4_5 \bigg{)}\\
\cline{1-3}
&&\\
\frac{1}{MM'} \bigg{(} \frac{1}{2} n_4^4 {p'}^g_5 - {p^4_5}^*
{p'}^g_4 
              \bigg{)}&
- \frac{1}{2M} (\rho - \Sigma\vert\mu^f\vert^2)& 
\frac{1}{2MM'} \bigg{(} n_4^4 {p'}^4_5 + {n'}^4_4 {p^4_5}^*
\bigg{)}\\
&
+ \frac{1}{2M} n^4_4 + \frac{M}{2{M'}^2} {n'}_5^5&
\\
&
- \frac{1}{M} {p^4_5}^* {p'}^5_4&
\\
- \frac{1}{M'} \bigg{(} {p^5_4}^* {p'}^g_5 + {p'}^g_4 \bigg{)}&
\frac{1}{2{M'}^2} {n'}^5_5 {p^5_4}^*&
- \frac{1}{2M'} (\rho' - \Sigma\vert\mu'_f\vert^2)\\
&
+ \frac{1}{2M^2} n^5_5 {p'}^5_4
&
+ \frac{M'}{2M^2} n^5_5 + \frac{1}{2M'} {n'}^4_4\\
&
&
- \frac{1}{M'} {p^5_4}^* {p'}^4_5
\end{array}
}\right)\,,
\\
\\
\end{array}
\end{equation}
where $\rho$ is defined above in  Appendix,
and $\rho'$ can be obtained from $\rho$ by the usual substitutions
$\mu^f\leftrightarrow {\mu'_f}^*$, 
$m_4\leftrightarrow m_4^*$, $m_5\leftrightarrow m_5^*$, 
$M\leftrightarrow M'$.

\end{document}